\documentclass[doublecol]{epl2}
\title{Pressure shift of the superconducting T$_c$ of LiFeAs}
\shorttitle{Title} %Insert here a short version of the title if it exceeds 70 characters

\author{Melissa Gooch\inst{1} \and Bing Lv\inst{2} \and Joshua H. Tapp\inst{2} \and Zhongjia Tang\inst{2} \and Bernd Lorenz\inst{1} \and Arnold M. Guloy\inst{2} \and Paul C.W. Chu\inst{1,3,4}}
\shortauthor{M. Gooch \etal}

\institute{
  \inst{1} TCSUH and Department of Physics, University of Houston, Houston, TX 77204, USA\\
  \inst{2} TCSUH and Department of Chemistry, University of Houston, Houston, TX 77204, USA\\
  \inst{3} Lawrence Berkeley National Laboratory, 1 Cyclotron Road, Berkeley, CA 94720, USA\\
  \inst{4} Hong Kong University of Science and Technology, Hong Kong, China
 }

\pacs{74.25.Dw}{Superconductivity phase diagrams} \pacs{74.62.Fj}{Pressure effects} \pacs{74.70.Dd}{Ternary, quarternary and multinary
compounds}

\abstract{The effect of hydrostatic pressure on the superconductivity in LiFeAs is investigated up to 1.8 GPa. The superconducting transition
temperature, T$_c$, decreases linearly with pressure at a rate of 1.5 K/GPa. The negative pressure coefficient of T$_c$ and the high ambient
pressure T$_c$ indicate that LiFeAs is the high-pressure analogue of the isoelectronic SrFe$_2$As$_2$ and BaFe$_2$As$_2$.}

\begin{document}

\maketitle

\section{Introduction}
The discovery of superconductivity in quaternary rare-earth transition-metal oxypnictides, R(O,F)FeAs (R=rare earth), with critical temperatures
up to 55 K \cite{kamihara:08,takahashi:08,chen:08b,chen:08,ren:08,ren:08b} has stimulated extensive research and discussions about the nature of
superconductivity in these compounds and possible similarities to the high-T$_c$ copper oxide superconductors. The layered structure and the
doping-induced (F replacing O) superconductivity as well as the existence of a magnetically ordered phase in the undoped (non-superconducting)
parent compounds indeed point toward similarities of pnictide and cuprate superconductors. While the Fe and As form a rigid metallic layer with
both atoms covalently bonded, the intermediate R(O,F) layer provides control of charges depending on the doping level. In the undoped ROFeAs
(also denoted as "1111" compound) the formal charge per FeAs is -1 (R$^{3+}$O$^{2-}$Fe$^{2+}$As$^{3-}$), it is metallic and shows a structural
and a spin density wave transition at lower temperatures \cite{delacruz:08}.

The active FeAs layer is structurally rigid and it can be chemically and structurally combined with other, oxygen-free charge reservoir layers.
For example, The RO layer in ROFeAs can be replaced by a single plane of alkaline earth atoms (Ae = Ca, Sr, Ba) forming the "122" structure,
AeFe$_2$As$_2$. The formal valence of Ae$^{2+}$[(FeAs)$^{1-}$]$_2$ indicates that the charge in the FeAs layer is the same as in ROFeAs, namely
-1 per FeAs pair. The physical properties of AeFe$_2$As$_2$ are similar to the "1111" compounds, in that they exhibit a simultaneous spin
density wave and a structural transition at a given temperature, for example T$_{S}$=205 K in SrFe$_2$As$_2$
\cite{rotter:08c,krellner:08,goldman:08}. Doping with K or Cs, replacing the Ae ion results in the appearance of superconductivity above a
critical concentration of alkali metals \cite{sasmal:08,chen:08d,rotter:08b}. It is interesting to note that even with the complete substitution
with K or Cs, the ternary compounds KFe$_2$As$_2$ and CsFe$_2$As$_2$ are still superconducting, albeit with low T$_c$'s below 4 K.

Extending the series of ternary alkali metal FeAs-compounds to smaller ionic radius it was shown that the "122" structure becomes unstable and
LiFeAs formed with a "111" composition with an unexpectedly high T$_c$=18 K \cite{wang:08,tapp:08,pitcher:08}. The structure of LiFeAs is the
unfilled version of the "1111" structure of the ternary ROFeAs. The formal charge balance, Li$^{1+}$Fe$^{2+}$As$^{3-}$, leaves an average charge
of -1 per FeAs pair, similar to the undoped, non-superconducting "1111" and "122" compounds. Considering that the undoped "1111" and "122"
compounds and LiFeAs have the same charge density in the FeAs layers, the existence of superconductivity in LiFeAs and the absence of any
magnetic order \cite{felner:08} appears puzzling and warrants further investigations. It has been shown recently that the application of
external pressure and its effect on the superconducting T$_c$ provides important information about the superconducting state of the FeAs-based
compounds \cite{gooch:08}. Furthermore pressure can even induce superconductivity in the undoped "122" compounds, AeFe$_2$As$_2$
\cite{alireza:08,fukuzawa:08,torikachvili:08}. We therefore investigated the pressure dependence of T$_c$ of LiFeAs and found that T$_c$(p)
decreases linearly with pressure at a rate of 1.5 K/GPa.

\section{Experimental}
Polycrystalline ceramic samples of LiFeAs were synthesized as described earlier \cite{tapp:08}. At ambient pressure the samples show a bulk
superconducting transition at 18 K. The ac magnetic susceptibility was measured under pressure through an inductance transformer. A dual coil
system was wrapped directly around the sample and the mutual inductance was measured using the low-frequency (19 Hz) LR700 Inductance Bridge
(Linear Research). A beryllium copper clamp cell was used to generate pressure up to 1.8 GPa. A 1:1 mixture of liquid Fluorinert FC70 and FC77
was used as the pressure transmitting medium. The pressure was determined in situ through the superconducting transition of high-purity lead
exposed to the same pressure next to the sample \cite{chu:74}.

\begin{figure}
\onefigure{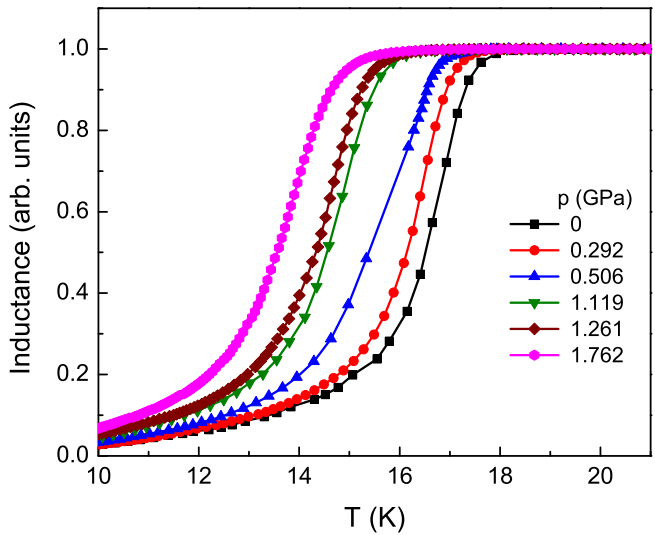} \caption{Normalized inductance of LiFeAs at different pressures.} \label{fig.1}
\end{figure}

\begin{figure}
\onefigure{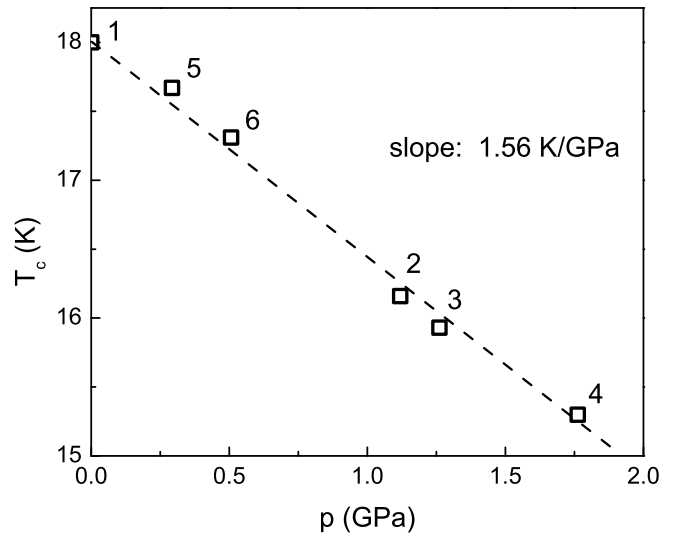} \caption{Pressure shift of T$_c$ of LiFeAs. The numbers next to the data points indicate the sequence of experiments
conducted.} \label{fig.2}
\end{figure}

\section{Results and Discussion}
The mutual inductance as measured for LiFeAs, at different pressures, is shown in Fig. 1. The data are normalized for better clarity so that the
inductance varies between 0 (low T limit) and 1 (at 25 K). The diamagnetic drop of the inductance below 18 K clearly marks the superconducting
transition. The relative sharpness of the transition indicates the high quality and phase uniformity of the sample. The inductance curves shift
nearly parallel to lower T with increasing pressure. After applying the highest pressure of this experiment (1.76 GPa) pressure was released to
verify that the compound is stable under the applied pressure. The data numbered "5" and "6" in Fig. 2 have been measured in a second pressure
cycle and they fit perfectly with the first pressure cycle. The T$_c$ of LiFeAs decreases linearly with pressure at a rate of 1.56 K/GPa, as
shown in Fig. 2.

The negative and linear pressure shift of T$_c$ has to be compared with the results of other FeAs-based superconductors. We have recently shown
that the sign of the pressure coefficient of T$_c$ depends on the doping state in FeAs superconductors \cite{lorenz:08}. This result was
supported by a systematic investigation of the pressure effect in the phase diagram of the hole-doped K$_x$Sr$_{1-x}$Fe$_2$As$_2$
\cite{gooch:08}. A linear decrease was only observed in the overdoped compound K$_{0.7}$Sr$_{0.3}$Fe$_2$As$_2$ with a large relative pressure
coefficient dlnT$_c$/dp=-0.18 GPa$^{-1}$ comparable with dlnT$_c$/dp=-0.09 GPa$^{-1}$ for LiFeAs. However, it is not conceivable to assume that
LiFeAs is in an overdoped state since the charge count of -1 per FeAs would rather locate it closer to the undoped FeAs compounds. This also
raises the question why LiFeAs is superconducting at all, even at ambient pressure.

The Li ion is the smallest ion in the alkali metal series which results in a significant compression of the LiFeAs structure as compared with
the rare earth "1111" or the alkaline earth "122" compounds. Some structural parameters of different compounds are listed in Table 1. Among all
FeAs based compounds LiFeAs has one of the smallest set of lattice parameters and volume (only CaFe$_2$As$_2$ has a smaller volume, but the c/a
ratio is significantly different from that of LiFeAs). Therefore, LiFeAs can be considered as a compressed structure where the chemical pressure
is due to the small ionic radius of Li. It appears conceivable to compare LiFeAs with other (undoped) FeAs compounds under high pressure
conditions. CaFe$_2$As$_2$, SrFe$_2$As$_2$ and BaFe$_2$As$_2$ have been found superconducting above a critical pressure of 0.23, 2.7 and 2.5
GPa, respectively \cite{torikachvili:08,alireza:08}. It is interesting that the superconductivity in these "122" compounds arises quickly above
the critical pressure but T$_c$ decreases continuously with further increasing p. We therefore propose that, at ambient pressure, LiFeAs is in a
chemically compressed state that is equivalent to the high-pressure state of the undoped "122" compounds above their critical pressure. The
negative dT$_c$/dp of LiFeAs would be consistent with this conclusion and it indicates that the T$_c$ of LiFeAs could be enhanced by a negative
pressure (or tensile strain). Tensile strain can be achieved by thin film deposition on a substrate with an appropriate lattice mismatch.

The above conclusion is further supported by the missing spin density wave state in LiFeAs. While all undoped "122" and "1111" compounds exhibit
the SDW transition at T$_S$ it was also shown that external physical pressure results in a decrease of T$_S$ and a suppression of the SDW state
\cite{lorenz:08,gooch:08,fukuzawa:08,kumar:08}. Because of the already compressed structure of LiFeAs the SDW phase is completely absent at
ambient pressure. This can also explain the relatively high superconducting T$_c$ despite the fact that the charge balance of -1 per FeAs is
typical for non-superconducting FeAs compounds.

\section{Summary and Conclusions}
The high ambient pressure superconducting T$_c$, the negative pressure coefficient dT$_c$/dp, and the missing spin density wave phase in LiFeAs
are explained by a chemical pressure effect compressing the lattice significantly even at zero physical pressure. LiFeAs is the high-pressure
analogue of the undoped FeAs-based compounds. The current high-pressure data suggest that higher T$_c$'s can be achieved if LiFeAs is deposited
as thin film on a substrate that generates a tensile strain.

\begin{table}
\caption{Comparison of structural parameters of LiFeAs with typical "122" and "1111" compounds.} \label{tab.1}
\begin{center}
\begin{tabular}{lccc}
 & a [${\AA}$] & c$^*$ [${\AA}$] & V$^*$ [${\AA}^3$] \\
\hline
LiFeAs \cite{tapp:08} & 3.7914 & 6.364 & 91.48 \\
SrFe$_2$As$_2$ \cite{sasmal:08} & 3.9261 & 6.138 & 95.38 \\
KFe$_2$As$_2$ \cite{sasmal:08} & 3.8414 & 6.919 & 102.10 \\
LaOFeAs \cite{delacruz:08} & 4.0301 & 8.737 & 141.90 \\
\end{tabular}
\end{center}
$^*$ For the "122" compounds c and V are divided by a factor of 2 for a better comparison with the "111" and "1111" compounds.
\end{table}

\acknowledgments This work is supported in part by the T.L.L. Temple Foundation, the J.J. and R. Moores Endowment, the State of Texas through
TCSUH, the USAF Office of Scientific Research, and at LBNL through USDOE. Support from the NSF (CHE-0616805) and the R.A. Welch Foundation is
gratefully acknowledged.

\bibliographystyle{eplbib}
%\bibliography{FeAs}

\end{document}